\documentclass{article}

\usepackage{PRIMEarxiv}
\usepackage{amssymb}
\usepackage[T1]{fontenc}
\usepackage{graphicx}
\usepackage{color}
\usepackage{color, soul}
\usepackage{microtype}
\soulregister\cite7
\soulregister\ref7
\soulregister\pageref7
\usepackage{multirow}
\usepackage{comment}
\usepackage{amssymb}
\usepackage{pifont}
\usepackage{amsmath} 
\usepackage{dsfont}
\usepackage{url}
\usepackage{tikz}
\usetikzlibrary{shapes.geometric, arrows}
\tikzstyle{box} = [rectangle, rounded corners, minimum width=0.2\linewidth, minimum height=1cm,text centered, text width=0.2\linewidth, draw=black, fill=white!30]

\tikzstyle{arrow} = [thick,->,>=stealth]

\title{CTI Dataset Construction from Telegram}

\author{
Dincy R. Arikkat \\
Department of Computer Applications, \\
Cochin University of Science \\
and Technology, India \\
\texttt{dincyrarikkat@cusat.ac.in} \\
\And
Sneha B. T. \\
Department of Computer Applications, \\
Cochin University of Science \\
and Technology, India \\
\texttt{snehabtsneha@pg.cusat.ac.in} \\
\And
Serena Nicolazzo \\
Department of Science and Technological Innovation, \\
University of Eastern Piedmont, \\
V.le Teresa Michel, 11, Alessandria, Italy\\
\texttt{serena.nicolazzo@uniupo.it} \\
\And
Antonino Nocera \\
Department of Electrical, Computer \\
and Biomedical Engineering, \\
University of Pavia, \\
A. Ferrata, 5, Pavia, Italy \\
\texttt{antonino.nocera@unipv.it} \\
\And
Vinod P. \\
Department of Computer Applications, \\
Cochin University of Science \\
and Technology, India \\
\texttt{vinod.p@cusat.ac.in} \\
\And    
Rafidha Rehiman K. A.\\
Department of Computer Applications, \\
Cochin University of Science \\
and Technology, India \\
\texttt{rafidharehimanka@cusat.ac.in} \\
\And
Karthika R.\\
Department of Computer Applications, \\
Cochin University of Science \\
and Technology, India \\
\texttt{karthikar@pg.cusat.ac.in} \\
}

\begin{document}

\maketitle

\date{September 2025}

\begin{abstract}
Cyber Threat Intelligence (CTI) enables organizations to anticipate, detect, and mitigate evolving cyber threats. Its effectiveness depends on high-quality datasets, which support model development, training, evaluation, and benchmarking. Building such datasets is crucial, as attack vectors and adversary tactics continually evolve. Recently, Telegram has gained prominence as a valuable CTI source, offering timely and diverse threat-related information that can help address these challenges. In this work, we address these challenges by presenting an end-to-end automated pipeline that systematically collects and filters threat-related content from Telegram. The pipeline identifies relevant Telegram channels and scrapes $145,349$ messages from $12$ curated channels out of $150$ identified sources. To accurately filter threat intelligence messages from generic content, we employ a BERT-based classifier, achieving an accuracy of $96.64\%$. From the filtered messages, we compile a dataset of $86,509$ malicious Indicators of Compromises, including domains, IPs, URLs, hashes, and CVEs. This approach not only produces a large-scale, high-fidelity CTI dataset but also establishes a foundation for future research and operational applications in cyber threat detection.

\end{abstract}


\section{Introduction}

Cyber Threat Intelligence (CTI) has become indispensable for security analysts, enabling them to identify, collect, manage, and disseminate information on vulnerabilities and attacks, and to respond proactively to emerging threats \cite{brown20212021}. Within the CTI lifecycle, data collection encompassing sources such as security alerts and threat intelligence reports from the web represents a critical foundational stage \cite{arazzi2025nlp}.

In this context, one challenge is that not all threat intelligence is published in standard CTI databases or integrated into commercial security platforms. Valuable CTI is often disseminated through unstructured channels such as blogs, social media posts, or reports from security companies and independent experts. To capture these dispersed insights, multiple online sources can be leveraged as early signals of emerging cyber threats. Information gathering thus becomes the first and most critical step, enabling the collection of relevant data on newly discovered vulnerabilities, active exploits, security alerts, threat intelligence reports, and security tool configurations. Curating CTI datasets requires addressing key challenges, including data sourcing from heterogeneous streams, ensuring data reliability, preserving privacy, and mitigating bias. A well-designed CTI dataset not only accelerates the advancement of automated threat intelligence systems but also strengthens global cyber defense capabilities through knowledge sharing and standardized evaluation frameworks. While platforms like Twitter \cite{shin2021twiti} have been widely explored for their CTI potential, other communication ecosystems remain underexamined. Among them, messaging applications, particularly Telegram\footnote{https://web.telegram.org}, have experienced exponential growth, evolving into key venues not only for general interaction but also for niche communities engaged in cybersecurity discourse, tool dissemination, and, in some cases, illicit activities. Telegram is a cloud-based messaging platform recognized for speed, privacy, and scalability in communication. It hosts public channels and groups covering a range of general and specialized topics, including cybersecurity and threat activity. Thanks to its openness and extensive global reach, Telegram has emerged as a significant source of Open Source Intelligence (OSINT), especially for tracking and analyzing emerging cyber threats, positioning it as a potentially valuable yet challenging source for CTI \cite{ravi2023exploring}.

Building on these considerations, this study presents a comprehensive dataset comprising $145,349$ messages collected from $12$ Telegram channels between January 2023 and February 2025. Since messages from identified CTI sources may include content unrelated to threat intelligence, we developed a filtering mechanism to identify relevant CTI messages based on transformer models. 
Such filtering is a critical preprocessing step, as it eliminates generic, non-security-related content and ensures that downstream models are trained exclusively on high-fidelity CTI data. Following the identification of relevant messages, we construct an IoC dataset from Telegram content for CTI analysis.


In summary, the key contributions of this work are as follows:

\begin{itemize}
\item We systematically identified $12$ high-value Telegram channels (from $150$ candidates) as reliable CTI sources and implemented a custom crawler to continuously collect intelligence-rich content.

\item We compiled a large-scale dataset of $145,349$ messages spanning two years, providing a substantial and timely resource for advancing CTI research.

\item We designed and thoroughly evaluated a BERT-based automated filtering model that achieved high accuracy in identifying cybersecurity-relevant intelligence, thereby ensuring the dataset’s reliability for downstream CTI applications.
\item We compiled Indicators of Compromise dataset that can support both research and operational cyber threat detection.
\end{itemize}

The rest of the article is organized as follows: Section~\ref{sec:related} reviews related work, Section~\ref{sec:approach} details our CTI dataset compilation, Section~\ref{sec:result} presents experiments and evaluation, and Section~\ref{sec:conclusion} concludes the paper.

\section{Related Work}
\label{sec:related}
A wide range of online platforms, including security blogs, forum posts, and Online Social Networks (OSNs), are frequently leveraged by both cybersecurity vendors and malicious actors to disseminate CTI in highly unstructured formats. These early disclosures often precede formal reporting and integration into authoritative and standardized repositories such as the Common Vulnerabilities and Exposures (CVE\footnote{\url{https://cve.mitre.org}}) database or the National Vulnerability Database (NVD\footnote{\url{https://nvd.nist.gov}})
\cite{threatt2024some}. 
While CVE and NVD provide timely and potentially critical insights but focus exclusively on known vulnerabilities. 
Hence, several researchers and practitioners have started to collect OSINT data through custom crawlers \cite{tabatabaei2017osint}. Crawling for CTI is not confined to Clear Web resources (the publicly accessible portion of the Internet) but also extends to the Dark Web, Deep Web, and OSNs \cite{arazzi2025nlp}.

Several domain-specific crawlers target the Clear Web. For example, inTime \cite{koloveas2019crawler} and MalCrawler \cite{singh2017malcrawler} are optimized to identify relevant pages before initiating the crawl, allowing them to filter out benign content and improve efficiency. 
Few contributions propose the construction of a CTI-relevant dataset \cite{zhou2023cdtier,wang2022aptner}, but specifically deal with NER and RE tasks within the CTI domain \cite{arikkat2024relation}. 
In the context of OSN-based crawling, several works~\cite{le2017sonar, alves2021processing, rodriguez2019generating, kristiansen2020cti, shin2021twiti} have focused on leveraging Twitter as a primary CTI source. These approaches are capable of detecting, geolocating, categorizing, and tracking cybersecurity-related events in real time by monitoring the Twitter stream. Typically, they rely on a curated list of seed keywords, often provided by domain experts, that serve as input to the streaming API, allowing the collection, detection, and classification of cyber threat indicators from tweets. However, other platforms such as Reddit, Pastebin, and GitHub have also been explored for CTI extraction~\cite{horawalavithana2019mentions,vahedi2021identifying}.
Although Telegram presents significant potential as a CTI source, systematic approaches for extracting intelligence from the platform remain limited \cite{ravi2023exploring}. Several obstacles complicate this process, including the overwhelming volume of messages, the frequent presence of non-English content, the unstructured and conversational style of discussions, and technical constraints such as API limitations that hinder large-scale data collection \cite{dutta2020overview, rahman2023attackers}. Overcoming these challenges requires automated methods capable of efficiently processing high-throughput message streams while filtering actionable intelligence from background noise. Initial efforts have begun to bridge this gap, combining AI-based models with human annotation to analyze Telegram-derived threats \cite{ravi2023exploring}, alongside broader research on automated CTI extraction from social data streams that can be adapted to this domain \cite{zhao2020timiner}. In addition, specialized tools such as TelegramScrap have emerged to address platform-specific scraping limitations, further supporting data acquisition.

\section{Proposed Approach}
\label{sec:approach}

\begin{figure*}
    \centering
    \includegraphics[scale=0.75]{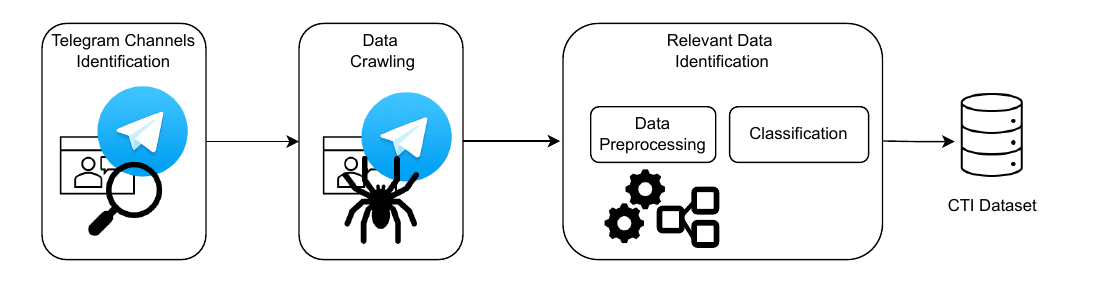}
    \caption{Architecture for the construction of the CTI dataset from Telegram}
    \label{fig:architecture}
\end{figure*}

The workflow begins with the \textit{selection of suitable Telegram channels} for data collection (Section~\ref{sub:selection}). Next, the \textit{gathering phase} is conducted through the platform’s standard interface (Section~\ref{sub:collection}). The collected messages then undergo \textit{text cleaning and preprocessing}, after which multiple \textit{BERT-based models} are trained to automatically filter valuable CTI content from irrelevant material (Section~\ref{sub:identification}). Finally, we compile an IoC dataset to support in-depth CTI analysis (Section~\ref{sub:ioc_extraction}).  An overview of the framework architecture is provided in Figure~\ref{fig:architecture}.

\subsection{Telegram Channels Identification}
\label{sub:selection}
In this phase, our objective is to identify Telegram channels that actively discuss attacks, threats, vulnerabilities, and share indicators of compromise (IoCs). To achieve this, we surveyed approximately 150 public channels, drawing from both prior research and open-source repositories. In particular, we referenced DarkGram \cite{roy2025darkgram}, as well as publicly available Telegram channel lists such as BreachSense’s catalog of threat actor channels\footnote{\url{https://www.breachsense.com/threat-actor-channels/}} and one GitHub repostory \footnote{\url{https://github.com/ghostwond3r/telegram_channel}}. In addition, we performed manual exploration within Telegram using cyber-related keywords (e.g., IoCs, CVE, DDoS, cyber attack, malware, ransomware, etc), thereby ensuring comprehensive coverage of channels relevant to CTI. Subsequently, we evaluated each channel against five criteria: {\em(i)} demonstrated relevance to threat intelligence \cite{zhao2020timiner}, {\em(ii)} depth of technical discussion and frequency of activity, {\em(iii)} primary language (English), {\em(iv)} evidence of direct IoC sharing, and {\em(v)} whether the channel was active and accessible. Following this multi-criteria assessment, we selected 12 channels deemed most suitable for in-depth data collection.

\begin{table*}[htbp]
\footnotesize
\caption{Cybersecurity Telegram Channels Statistics}
\centering
\begin{tabular}{|c|p{4.5cm}|c|c|p{0.8cm}|p{5cm}|}
\hline
\textbf{Number} & \textbf{Channel Name} & \textbf{Messages} & \textbf{Subscribers} & \textbf{AMD*} & \textbf{Top 5 words}\\
\hline
C1 &DLM - CVE Monitor & 19,142 & 884 & 84 & cve, vulnerability, affected, link, products
 \\
\hline
 C2& Cybersecurity \& Privacy - News & 27,254 & 23,542 & 38 & vulnerability, cve, database, cibsecurity, security
\\
\hline
 C3 & Pro-Palestine Hackers Movement & 967 & 5,295 & 6 & company, data, website, israeli, attacked
\\
\hline
 C4 & Z-BL4CK-H4T & 250 & 4,359 &17 & israel, website, rippersec, investigation, undergroundnet
\\
\hline
C5 & RipperSec & 4,510 & 5,151 & 26 & rippersec, sedihcrew, zenimous, target, team
\\
\hline
 C6 & Dark Web Informer - Cyber Threat Intelligence - CVE Alerts & 6,021 & 157 & 118 & cve, threat, intelligence, cyber, vulnerability
\\
\hline
 C7& BleepingComputer & 2,489 & 8,544 & 6 & data, windows, security, ransomware, microsoft
\\
\hline
 C8 & The Hacker News & 3,472 & 144,734 & 5 & security, malware, data, cve, critical
\\
\hline
 C9& CVE Notify & 32,283 & 15,285 & 114 & cve, vulnerability, issue, user, attacker
\\
\hline
C10& CVE Tracker & 16,227 & 64 & 87 & cve, vulnerability, affected, link, products
\\
\hline
C11& Cyber Threat Intelligence & 30,927 & 31,267 & 86 & cyber, cve, security, attack, data
\\
\hline
C12&  Hackmanac Cyber Alerts & 1,807 & 2,738 & 9 & data, group, cyberattack, ransomware, alert
\\
\hline
& \textbf{Total} & \textbf{145,349} & \textbf{242,020} & \textbf{186}& cve, vulnerability, affected, cvss, link\\
\hline
\end{tabular}
\label{tab:channels}
\textbf{*AMD - Average Message per Day}
\end{table*}

\subsection{Data Crawling}
\label{sub:collection}

\begin{figure}
    \centering
    \includegraphics[scale=0.28]{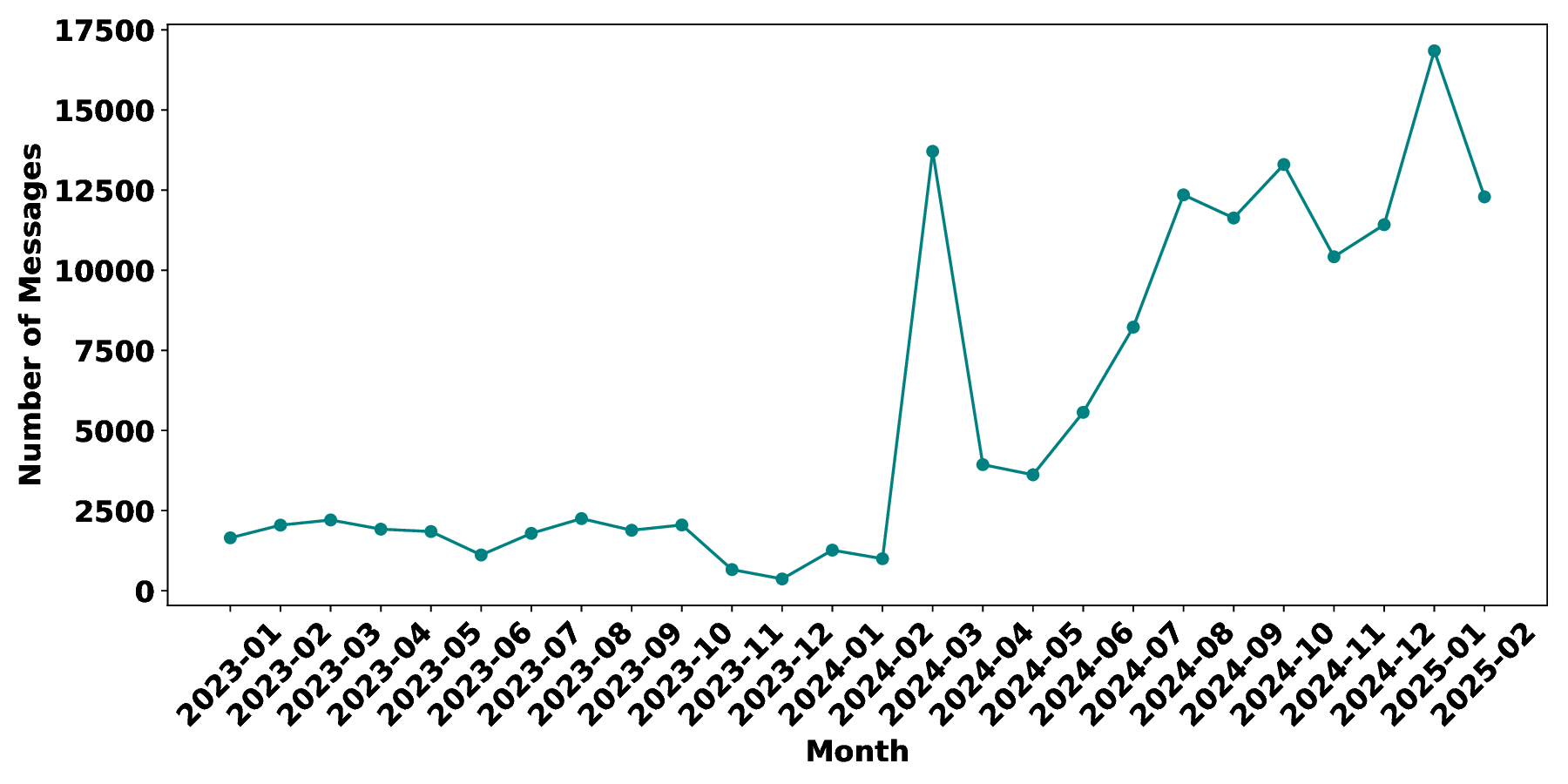}
    \caption{Monthly Message Volume Across All Telegram Channels}
    \label{fig:month_statistics}
\end{figure}

After identifying the target channels as trusted data sources, we proceeded with the crawling phase. For this task, we relied on the official Telegram API and, in particular, the Telethon library\footnote{[https://docs.telethon.dev/en/stable]} (asynchronous MTProto API client for Python). 
Using this setup, we scraped messages from the selected channels between January 2023 and February 2025, resulting in a dataset of $145,349$ messages. Table~\ref{tab:channels} summarizes the selected channels along with their respective message volumes, subscriber counts, average message per day, and top five words. Figure \ref{fig:month_statistics} illustrates the monthly distribution of posts across all channels. From the figure, it is evident that the volume of messages posted during 2024–2025 was substantially higher compared to 2023.

\subsection{Relevant CTI Data Classification}
\label{sub:identification}
While our corpus comprises $145,349$ messages crawled from CTI-related Telegram channels, not all messages explicitly reference attack incidents, vulnerability, or IoCs. To address this limitation, we introduce an additional filtering stage designed to retain only content that is directly relevant to CTI. Specifically, we develop a classification model to distinguish CTI-relevant messages from unrelated chatter. Before proceeding with the classification tasks, we perform a preprocessing step that reduces noise and produces cleaner text for downstream analysis. This preprocessing stage consists of a sequence of operations: {\em(i)} \textbf{IoC Normalization:} CTI messages often contain IoCs such as IP addresses, URLs, CVE identifiers, and file hashes, which may appear in standard, defanged, or obfuscated forms. We normalized these entities by replacing them with placeholder tokens: \texttt{[ip]}, \texttt{[url]}, \texttt{[cve]}, and \texttt{[hash]} to retain key threat information. 
{\em(ii)} \textbf{Lowercasing:} 
    All text is converted to lowercase to ensure uniform representation, reducing vocabulary sparsity and simplifying downstream processing.
{\em(iii)} \textbf{Removal of emojis and non-essential special characters:} While emojis are commonly used on social media to convey user emotions or sentiment, they typically introduce noise rather than substantive content for identifying CTI relevance. Consequently, they were removed to reduce data sparsity and also to remove special characters.
{\em(iv)} \textbf{Lemmatization:} 
   We use lemmatization to reduce words to their base form, allowing the model to generalize across related variants (e.g., \textit{attacks}, \textit{attacking}, \textit{attacked} $\rightarrow$ \textit{attack}).

Subsequently, we leveraged the transformer-based models to identify and filter CTI-relevant content from our collected corpus. Specifically, we employed models including {\em (i)} standard BERT (Bidirectional Encoder Representations from Transformers) \cite{lee2018pre}, {\em (ii)} DistilBERT (Distilled version of BERT) \cite{sanh2019distilbert}, {\em (iii)} RoBERTa (Robustly Optimized BERT) \cite{liu2019roberta}, {\em (iv)} CySecBERT \cite{bayer2022cysecbert},  and {\em (v)} SecBERT\footnote{\url{https://github.com/jackaduma/SecBERT}}, which are widely used in Natural Language Processing tasks \cite{sun2019fine}.

\subsection{IoC Extraction and Verification}
\label{sub:ioc_extraction}
To extract potential threat indicators from Telegram messages, we developed a set of tailored regular expressions (RegEx). These expressions were designed to capture web URLs, IP addresses, domain names, file hashes, and CVE identifiers. 
Since RegEx-based extraction may also capture benign or irrelevant entries, we performed additional validation. Indicators were cross-checked using VirusTotal\footnote{\url{https://www.virustotal.com/gui/home/upload}}, and the NVD for CVE verification. Through this two-step enrichment process, we curated a refined set of malicious IoCs suitable for threat intelligence applications.

\section{Experiments}
\label{sec:result}
In this section, we evaluate the performance of five BERT-based models in identifying relevant CTI texts and also investigate the IoCs collected from the relevant messages. Model effectiveness is assessed using Accuracy and F1-score. To construct a reliable dataset for training the relevance classification model, we estimated the required sample size using the standard statistical approach for finite populations \cite{ahmad2017determining}. Based on a $95\%$ confidence level, a $1\%$ margin of error, and an assumed population proportion of $50\%$, the resulting sample size was approximately $9,009$ messages from the total corpus of $145,349$. The selected messages were manually annotated to create the labeled dataset. Each message was independently reviewed and assigned a label of either \textit{Relevant} (containing actionable threat intelligence) or \textit{Irrelevant} (lacking such content). To ensure annotation reliability, we measured inter-annotator agreement using Cohen’s Kappa ($k$) \cite{kilicc2015kappa}, which yielded $0.90$, indicating ``almost perfect” agreement and confirming the reliability of annotation. The annotation process, however, revealed an imbalance between \textit{Relevant} and \textit{Irrelevant} classes. To avoid classifier bias, we applied random under-sampling, yielding a balanced dataset of $8,634$ messages ($4,317$ Relevant, $4,317$ Irrelevant), which served as ground truth for training and evaluation. 
For model development, the dataset was divided into three subsets: $70\%$ was allocated for training, $10\%$ was used as a validation set, and the remaining $20\%$ was reserved as a test set. Table \ref{tab:performance} presents the comparative performance of the five evaluated models: BERT, DistilBERT, RoBERTa, CySecBERT, and SecBERT. Among these, the standard \texttt{bert-base-uncased} model achieved the strongest performance on our test data, attaining an Accuracy of $96.6\%$ and an F1-score of $0.97$. This high performance highlights the robustness and reliability of the BERT model in accurately identifying threat intelligence content. 

\begin{table}
\scriptsize
\centering
\caption{Performance Comparison of BERT-based Models}
\label{tab:performance}
\begin{tabular}{|p{2.5cm}|p{1.5cm}|p{2.5cm}|}
\hline
\textbf{Model} & \textbf{Accuracy} & \textbf{F1-Score (Class 0,1)} \\
\hline
DistilBERT & 95.83\% & 0.96, 0.96 \\
CySecBERT & 95.42\% & 0.95, 0.95 \\
RoBERTa & 96.00\% & 0.96, 0.96 \\
SecBERT & 95.19\% & 0.95, 0.95 \\
\textbf{BERT } & \textbf{96.64\%} & \textbf{0.97, 0.97} \\
\hline
\end{tabular}
\end{table}

After developing the relevant CTI content classification model, we applied it to the remaining unlabeled messages in the corpus. These messages underwent the same preprocessing steps as the training data and were then classified using the fine-tuned binary BERT model. As a result, the BERT model classified the messages into $99,340$ relevant and $42,510$ irrelevant messages. 

The relevant messages were further analyzed for IoC extraction to support threat intelligence. Using regular expressions, we initially extracted a total of $188,290$ indicators from the dataset, as summarized in Table~\ref{tab:channel_list}. Since not all extracted indicators were malicious, we performed verification using VirusTotal and the NVD database, which resulted in a refined set of $86,509$ confirmed malicious indicators (see Table~\ref{tab:channel_list}). The analysis of this curated collection revealed a notable distribution across different IoC types. Out of the total indicators collected, the majority were CVEs, followed by URLs, IPs, Domains, and Hashes. Specifically, CVEs accounted for about $45.5\%$ of all collected indicators, URLs $50.9\%$, IPs $2.1\%$, Domains $1.0\%$, and Hashes $0.5\%$. When considering only the malicious indicators, nearly all were CVEs, while URLs, IPs, Domains, and Hashes contributed only a small fraction. This threat indicator dataset serves as a benchmark for further threat analysis.
\begin{table}[htbp]
\small
\centering
\caption{IoCs extracted and validated per channel}
\label{tab:channel_list}

\begin{tabular}{|c|cc|cc|cc|cc|cc|}
\hline
\textbf{Channel} & \multicolumn{2}{|c|}{\textbf{Domain}} & \multicolumn{2}{|c|}{\textbf{IP}} & \multicolumn{2}{|c|}{\textbf{URL}} & \multicolumn{2}{|c|}{\textbf{Hash}} & \multicolumn{2}{|c|}{\textbf{CVE}} \\ \cline{2-11} 
 & \textbf{T} & \textbf{M} & \textbf{T} & \textbf{M} & \textbf{Total} & \textbf{M} & \textbf{T} & \textbf{M} & \textbf{T} & \textbf{M} \\ \hline
C1  & 237 & 19 & 731 & 25 & 24    & 1   & 173 & 0 & 19153 & 19145 \\ \hline
C2  & 354 & 36 & 605 & 22 & 26935 & 2   & 341 & 0 & 15827 & 15827 \\ \hline
C3  & 419 & 4  & 10  & 0  & 315   & 13  & 0   & 0 & 0     & 0     \\ \hline
C4  & 105 & 3  & 2   & 0  & 186   & 24  & 2   & 0 & 0     & 0     \\ \hline
C5  & 65  & 3  & 630 & 77 & 0     & 0   & 0   & 0 & 1     & 1     \\ \hline
C6  & 86  & 5  & 276 & 11 & 11476 & 160 & 30  & 1 & 5348  & 5348  \\ \hline
C7  & 21  & 8  & 2   & 0  & 4972  & 125 & 0   & 0 & 55    & 54    \\ \hline
C8  & 0   & 0  & 4   & 1  & 3452  & 17  & 0   & 0 & 475   & 474   \\ \hline
C9  & 277 & 24 & 923 & 38 & 18617 & 13  & 182 & 0 & 25503 & 25503 \\ \hline
C10 & 208 & 16 & 640 & 23 & 191   & 2   & 151 & 0 & 16309 & 16309 \\ \hline
C11 & 185 & 7  & 74  & 3  & 28977 & 51  & 0   & 0 & 3074  & 3073  \\ \hline
C12 & 7   & 0  & 0   & 0  & 628   & 9   & 0   & 0 & 32    & 32    \\ \hline
\end{tabular}
\vspace{1em}\\
*\textbf{T} - Total indicators extracted, \textbf{M} - Malicious indicators
\end{table}

\section{Conclusion}
\label{sec:conclusion}

In this work, we present a large-scale CTI dataset comprising $145,349$ messages collected from $12$ selected Telegram channels between January 2023 and February 2025. We train and evaluate a binary BERT-based classifier, designed to automatically filter cybersecurity-relevant messages with a high accuracy of $96.64\%$. From the curated messages, we assembled a comprehensive dataset of $86,509$ malicious IoCs. Our contributions include the systematic identification of high-value CTI channels on Telegram, the creation of a CTI message dataset with IoCs, and the development of an automated filtering pipeline that enhances the quality and usability of CTI data for research and operational purposes. In the future, we plan to expand this dataset, considering also blogs in the Dark/Deep Web and other Social Network scenarios. Moreover, we plan to evaluate IoCs using other threat intelligence feeds such as AlienVault, MalwareBazaar, etc., and also extract information about attack behaviours.

\bibliographystyle{splncs04}

\end{document}